\documentclass[useAMS,usenatbib]{mn2e}
\usepackage{psfig}
\input abbreviations.def
\input journals.def

\title[Star Cluster Ecology VII]{Star Cluster Ecology: VII The evolution of 
       young dense star clusters containing primordial binaries}

\author[Simon F.\ Portegies Zwart et al.]
        {
	Simon F.\ Portegies Zwart,$^{1,2}$\thanks{E-mail:
	spz@science.uva.nl; KNAW fellow} 
	Stephen L.\ W.\ McMillan$^{3}$ 
	Junichiro Makino$^4$  \\
$^{1,2}$    Astronomical Institute 'Anton Pannekoek',
	University of Amsterdam, Kruislaan 403, 
	1098SJ Amsterdam, the Netherlands, \\
	Section Computational Science,
	University of Amsterdam, Kruislaan 403, 
	1098SJ Amsterdam, the Netherlands \\
$^3$ 	Department of Physics,
	Drexel University,
        Philadelphia, PA 19104, USA \\
$^4$    Department of Astronomy, University of Tokyo, Tokyo 113, Japan \\
        }

\begin{document}
\maketitle

\date{Accepted 1643 December 32. Received 1687 December -1; 
      in original form 1687 October 3.6}

\pagerange{\pageref{firstpage}--\pageref{lastpage}} \pubyear{2004}

\label{firstpage}

\begin{abstract}
We study the first $\sim$100\,Myr of the evolution of isolated star
clusters initially containing 144179 stars, including 13107 (10\%)
primordial hard binaries.  Our calculations include the effects of
both stellar and binary evolution. Gravitational interactions among
the stars are computed by direct N-body integration using high
precision GRAPE-6 hardware.
The evolution of the core radii and central concentrations of our
simulated clusters are compared with the observed sample of young
($\aplt 100$\,Myr) star clusters in the large Magellanic cloud.
Even though our simulations start with a rich population of primordial
binaries, core collapse during the early phase of the cluster
evolution is not prevented. Throughout the simulations, the fraction
of binaries remains roughly constant ($\sim 10$\,\%).
Due to the effects of mass segregation the mass function of
intermediate-mass main-sequence stars becomes as flat as $\alpha=-1.8$
in the central part of the cluster (where the initial Salpeter mass
function had $\alpha=-2.35$).  About 6--12\% of the neutron stars were
retained in our simulations; the fraction of retained black holes is
40--70\%. In each simulation about three neutron stars become members
of close binaries with a main-sequence companion.  Such a binary will
eventually become an x-ray binary, when the main-sequence star starts
to fill its Roche lobe.  Black holes are found more frequently in
binaries; in each simulated cluster we find $\sim11$ potential x-ray
binaries containing a black hole.
Binaries consisting of two white dwarfs are quite common, but few
(20--30) are sufficiently close that they will merge within a Hubble
time due to the emission of gravitational radiation.  Clusters with
shorter relaxation times tend to produce fewer merging white dwarf
binaries.  The white dwarf binaries that do merge are all sufficiently
massive to produce a type Ia supernova.
The densest cluster produces about twice as many blue stragglers as a
field population containing the same number of binaries, and these
blue stragglers are more massive, bluer and brighter than in less
dense clusters.

\end{abstract}

\section{Introduction}

High-quality ground- and space-based observations over the past two
decades have revealed the existence of numerous young, dense star
clusters in our Galaxy and beyond.  Examples include (1) the Arches
\citep{2002ApJ...581..258F,2005ApJ...628L.113S} and Quintuplet
\citep{1999ApJ...514..202F} systems in the Galactic center
\citep{2003ApJ...594..812G,2003ApJ...586L.127G}; (2) the rich clusters
NGC 3603 \citep{1994ApJ...436..183M,2004AJ....128.2854M} and
Westerlund 1 \citep{1998A&AS..127..423P,2005A&A...434..949C} in the
Galactic disk; (3) the R136 \citep{1998ApJ...493..180M} system and
other young clusters in the Large Magellanic Cloud (LMC)
\citep{2003MNRAS.338...85M}; (4) an increasing number of young star
clusters in nearby starburst systems such as the Antennae and a newly
discovered cluster
\citep{2003A&A...397..177B,2003A&A...404..223B,2006ApJ...643.1166F}.

These systems are of great interest for a number of reasons.  First,
as both observations and simulation techniques continue to improve,
evolutionary studies of model systems allow us to probe the dynamical
state of observed clusters, and may offer key insights into the
conditions under which star clusters are born.  Second, these dense
clusters are likely to be the sites of complex physical phenomena,
such as stellar collisions and mergers
\citep{1999A&A...348..117P,2002ApJ...576..899P,2004ApJ...604..632G,2006MNRAS.368..141F},
placing them at the interface of stellar dynamics, stellar and binary
evolution, and stellar hydrodynamics.  Partly as a result of this
overlap of traditionally distinct astrophysical disciplines, the past
few years have seen an upsurge in interest in modeling dense stellar
systems, which pose significant theoretical and technical challenges
to researchers
\citep{2003NewA....8..337H,2003NewA....8..605S}.\footnote{See for
example {\tt http://manybody.org/modest}.}  Finally, since such
clusters may plausibly be the progenitors of globular-cluster like
systems, the studies presented here also offer valuable clues to the
early evolution of the globular cluster systems observed in many
galaxies.  The early evolutionary conditions considered here may also
have important consequences for the present-day content of globulars.

In performing simulations of young star clusters we run into an
immediate problem.  The initial conditions of these systems have been
actively debated for many years, but no consensus has been reached.
Models of star formation are as yet insufficiently advanced to provide
definitive predictions of initial structure for large systems
\citep{2001ApJ...556..837K,2002ApJ...576..870P,2005MNRAS.356.1201B},
and we cannot simply run an observed cluster backwards in time, even
if its parameters were all known to arbitrary accuracy.  Rather, we
start with a poorly determined but plausible initial state, evolve it
forward in time, then attempt to match observable properties of our
model cluster with actual clusters in the universe to assess the
reasonableness of our initial choice.

In this study we simulate young (age $\aplt 100$\,Myr) star clusters
by integrating the equations of motion of all stars and binaries.  We
use the Starlab environment (Portegies Zwart et al
2001),\nocite{2001MNRAS.321..199P} which acquires it greatest speed on
the GRAPE-6 special-purpose computer (GRAvity PipE, Makino et al 1997;
2003).\nocite{1997ApJ...480..432M,2003PASJ...55.1163M} The
calculations presented here were performed on the GRAPE hardware at
the University of Tokyo, the MoDeStA\footnote{See {\tt
http://modesta.science.uva.nl}} platform in Amsterdam, and the GRAPE-6
system at Drexel University.  Both stellar and binary evolution are
included self-consistently in our models.

\section{Simulations}\label{Sect:Simulation}

We focus on four basic cluster simulations (\#1--4), with initial
conditions summarized in Table\,\ref{Tab:initials}.  For each
realization, one full calculation from zero age to about 100\,Myr was
performed; for simulations \#1 and \#2 one additional realization was
computed.  As discussed in more detail below, these simulations differ
only in the choice of length (and hence time) scales, effectively
exploring the dependence of the evolution on the ratio of the cluster
relaxation time to the (fixed) time scale for stellar evolution.  In
each case, further simulations have been performed to test the
sensitivity of portions of the evolution to different choices of
physical parameters, such as the suppression of binary heating or
stellar mass loss.

\begin{table*}
\caption[]{Parameters of the four simulations.  In each case the total
  mass is $M=433209\,\msun$, the number of stars is 131072 (128k), the
  core mass is $M_{\rm core} = 0.00298M = 1282\,\msun$, and the core
  number is $N_{core} = 360$.  The runs differ only in the choice of
  virial radius, and therefore in the density within the virial radius
  ($\rho_{\rm vir}$). All contain an initial binary fraction of 10\%,
  with orbital parameters as described in the text.  }
\bigskip
\begin{tabular}{lrrcrrrr}
Simulation& \rvir& \rcore& $\rho_{\rm vir}$& \tcross & \trlx & \tcc \\
          &  [pc]& [pc]  & [\msun/pc$^3$]  & [Myr] & [Myr] \\
\hline
\#1	  &  1.27 & 0.010&   40100         & 0.032& 80   & 40  \\ 
\#2       &  3.2  & 0.026&    2500         & 0.129& 320  & 77  \\ 
\#3       &  8.1  & 0.066&     155         & 0.516& 1300 & $\apgt 100$ \\ 
\#4       & 20    & 0.162&      10.2       & 2.067& 5100 & $\apgt 100$ \\ 
\label{Tab:initials}
\label{Tab:results}
\end{tabular}
\end{table*}

The model clusters are initialized by selecting the number of stars,
the stellar mass function, the binary fraction and the distribution of
binary orbital elements, and the density profile.  For our most
compact model (simulation \#1) we adopt the initial conditions derived
by Portegies Zwart et al.\,(2004) \nocite{2004Natur.428..724P} to
mimic the 7--12\,Myr old star cluster MGG-11 in the starburst galaxy
M82, which was observed in detail by McCrady et
al.~(2003).\nocite{2003ApJ...596..240M} In this paper, however, we
extend the evolution of that model to about 100\,Myr.  The initial
half-mass relaxation time of simulation \#1 is 80 Myr; the other
simulations are performed with larger cluster radii, resulting in
longer relaxation times.

We summarize here the choice of initial conditions for simulation \#1.
We select $N=131072$ (128k) stars from a King
(1966)\nocite{1966AJ.....71...64K} $\Wo=12$ density distribution (see
Portegies Zwart et al.~2004).  For simplicity, stellar masses are
drawn from a Salpeter initial mass function ($\alpha = -2.35$) between
1\,\msun\, and 100\,\msun.  The lower limit is set by the recent
indications that mass functions in such young and massive star
clusters may be truncated below about 1\,\msun\, at least for several
star clusters including knot F in M82 \citep{2001MNRAS.326.1027S},
MGG-11 in the same galaxy \citep{2003ApJ...596..240M}, and the Arches
cluster in the Galactic center \citep{2005ApJ...628L.113S}.  The mass
function also appears to be truncated at the upper end, at around
100--150\,\msun
\citep{2004MNRAS.348..187W,2005Natur.434..192F}. Between these limits
the power-law can be described adequately by the Salpeter slope
\citep{1955ApJ...121..161S,2001MNRAS.322..231K}.  

The most widely used mass functions, such as
\cite{2001MNRAS.322..231K}, peak around 0.5\,\msun. If we had included
the low-mass stars from these distributions, the total number of stars
in our simulations would have increased by about a factor of 2.6,
while the total mass would have increased by about 30\%. The increased
relaxation time due to this change might have a significant effect on
the long-term dynamical evolution of the simulated clusters, but
during the first $\sim 100$\,Myr the effect is expected to be small,
since the early evolution of these clusters is dominated by the
massive stars. We therefore expect that the neglect of low-mass
($<1$\,\msun) stars does not profoundly affect our results.

There is no primordial mass segregation---that is, a star's initial
position is uncorrelated with its mass.  Ten percent of the stars are
selected randomly as binary primaries and are provided with a
companion (secondary) star with mass distributed uniformly between
1\,\msun\, and the mass of the primary.  The total mass of the cluster
is then $M \simeq 433000\,\msun$.  Binary parameters are determined by
choosing a binding energy and orbital eccentricity.  The latter is
taken randomly from a thermal distribution [$f(e)=2e$], whereas the
binding energy is taken randomly (uniform in $\log E$) between $E =
10$\,kT (corresponding to a separation of about 1000\,\rsun) and
maximum binding energy such that the corresponding distance between
the two stars at pericenter exceeds the sum of their radii.  (The
energy scale $kT$ is defined by the condition that the total stellar
kinetic energy of the system, excluding internal binary motion, is
$\frac32NkT$.)

For the other simulations (\#2--4), we adopt the same realization of
the initial stellar masses, positions and velocities (in virial N-body
units [Heggie \& Mathieu 1986]\nocite{HM1986}), but with a different
size and time scaling for the stellar evolution, such that the
half-mass two-body relaxation time (\trlx) for simulations \#2, \#3,
and \#4 are, respectively, 4, 16, and 64 times that for simulation
\#1.  The binary populations in these simulations therefore have
larger maximum orbital separations in clusters with longer relaxation
times, because the adopted minimum binding energy of $10\,kT$ shifts
to smaller physical values.  Simulations \#2, \# 3 and \# 4 have
maximum orbital separations of roughly 2000\,\rsun, 5000\,\rsun\, and
$10^4$\,\rsun, respectively.

After initialization we solve the equations of motion for the stars in
the cluster potential using the Starlab {\tt kira} integrator,
simultaneously calculating the evolution of the stars and binaries.
The stellar evolution model adopted is based on Eggleton, Fitchet \&
Tout (1989), \nocite{1989ApJ...347..998E} and the binaries are evolved
using {\SeBa} \citep{1996A&A...309..179P,1998A&A...332..173P}.  We
ignore any external tidal field, but stars are removed from the
simulation when they reach more than 60 initial half-mass radii from
the density center of the cluster.  The neglect of the tidal field
limits the validity of our results to relatively isolated clusters
like NGC\,3603 and Westerlund~1.  These initial conditions may also be
applicable to the star clusters in the LMC, like R\,136, as the low
density and irregular shape of this galaxy imposes only a shallow
background potential on these clusters. All calculations are continued
to an age of about 100\,Myr.

During the integration of simulation \#2, total energy is conserved on
average to better than one part in $10^{7.1\pm 1.7}$ per crossing
time, with a total fractional difference between the final and initial
energies (corrected for mass loss and other explicitly
non-conservative events, such as supernova explosions or stellar
collisions) of $\sim 10^{-4}$.  For the other runs the energy
conservation is at least an order of magnitude better.  Simulations
\#3 and \#4 have better energy conservation because they are larger
clusters for which fewer integration timesteps had to be taken and
during which fewer strong dynamical multibody encounters occurred.
Simulation \#1, although more compact than \#2, exhibited an early
phase of core collapse during which a collision runaway occurred (see
\cite*{2004Natur.428..724P}). Such events are generally easier for the
$N$-body integrator to handle, compared to the many strong dynamical
multibody interactions in simulation \#2.  In addition we performed a
second simulations with the same realization \#1, this second
simulation we call \#1R.  The main difference between simulation \#1
and \#1R is the treatment of supernovae in the massive collision
runaways occurring in those simulations.  In simulation \#1 the very
massive star collapses to a 39\,\msun\, black hole, losing about
1193\,\msun\, in the supernova, whereas in simulation \#1R the same
star collapses to a 1232\,\msun\, black hole without losing any mass
in the supernova explosion.

\section{Results}

In this section we discuss the early evolution of star clusters
\#1--4.  We focus on the observational consequences for these
simulations, discussing the stellar and binary populations.

\subsection{Evolution of cluster structural parameters}
 

\begin{figure}
\psfig{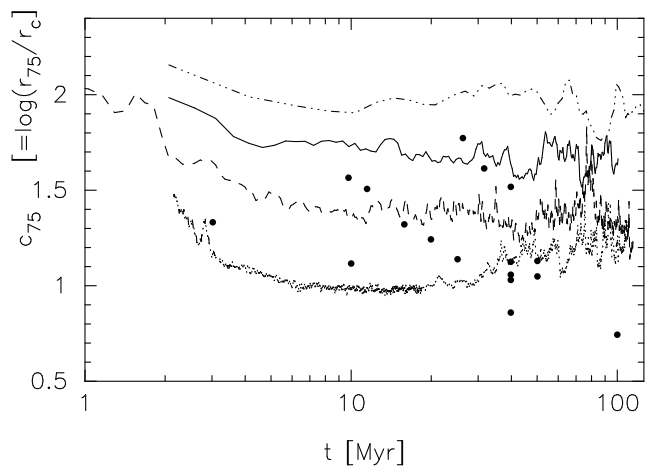}
   \caption[]{Evolution of the concentration $c_{75}$ [defined here as
   $\log_{10}(r_{75\%}/\rcore)$] for the simulated clusters \#1
   (dotted curve), \#2 (dashes), \#3 (solid) and \#4 (dash-3-dotted
   curve).  For comparison we show (bullets) the measured
   concentrations for the young LMC clusters from Mackey \& Gilmore
   (2003). 
   \label{fig:c}
   }
\end{figure}

\begin{figure}
   \psfig{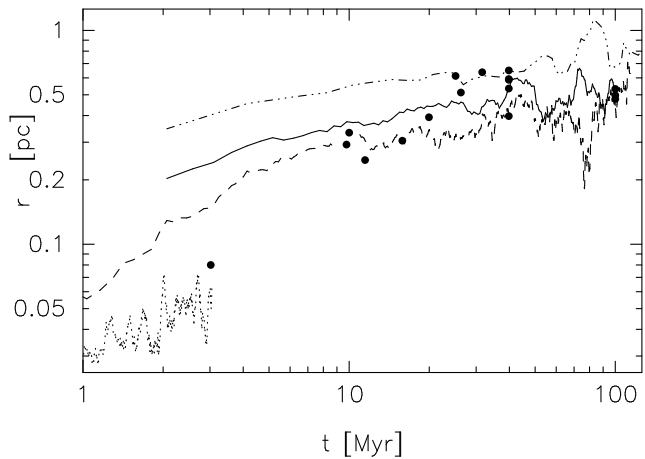}
   \caption[]{Measurement for the core radius (estimated as $0.25r_c$,
   with $r_c$ taken from \cite*{2003MNRAS.338...85M}) for the clusters
   listed in \cite{2003MNRAS.338...85M}.  To guide the eye we plot the
   evolution of the core radius of model \#1 (dotted curve), \#2
   (dashed line), \#3 (solid curve) and \#4 (dash-3-dotted curve).
   Note the unusually small core radius of the star cluster R136
   (leftmost point) which actually lies on the evolution of the core
   radius for simulation \#1 (not shown here).  The curve for
   simulation \#1 is only shown for the first few Myr, as at later
   times it becomes quite nosiy and runs into the curve for simulation
   \#2, making them both hard to distinguish.
\label{fig:rcore} }
\end{figure}

Figures \ref{fig:c} and \,\ref{fig:rcore} show the evolution of the
central concentration and core radii of our simulations. In addition,
we overplot as bullets the structure measurements for the LMC star
clusters reported by Mackay \& Gilmore (2003).  Those authors list a
total of 16 LMC clusters younger than 100\,Myr, and 25 younger than 1
Gyr. They provide some structural data, although not directly related
to the commonly used concentration parameter $c$, defined as $c =
\log\rt/r_{\rm c}$.  Since our simulation models are isolated, we also
have opted for a slightly modified definition of the concentration: we
use the 75\% Lagrangian radius ($r_{75}$) instead of the tidal radius
in the definition, and write $c_{75} \equiv \log r_{75}/r_{\rm c}$.
\cite{2003MNRAS.338...85M} do not measure the tidal radii of their
clusters, but from their structure parameters we can readily estimate
the 75\% Lagrangian radii.

\begin{figure}
   \psfig{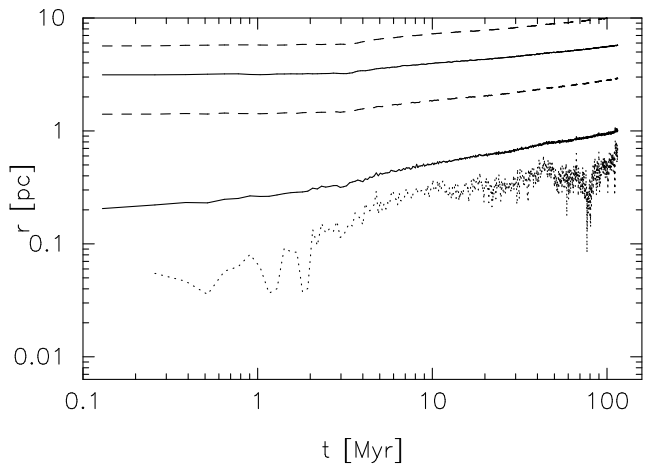}
   \caption[]{Evolution of the core radius (dotted curves), 5\%
    Lagrangian radius (lower solid), 25\% (lower dashed line), half
    mass radius (upper solid) and 75\% Lagrangian radii for
    simulation \#2.
   \label{fig:Lagrangian}
   }
\end{figure}

Figure \ref{fig:c} compares the concentration parameter ($c_{75}$)
derived from observations with the results of our simulations \#1--4,
which generally bracket the observed parameters except at very low
concentrations around 100\,Myr.  The low concentrations of these older
clusters may be due to a rather flat initial mass function which
drives a more dramatic expansion even at later times, by stellar-wind
mass loss \citep{2000ApJ...535..759T}, by dynamical heating due to the
presence of a large number of stellar mass black holes of the cluster
\citep{2004ApJ...608L..25M} or by a (much) larger population of
primordial binaries \citep{2006astro.ph..2408H,2006astro.ph..2409T}
than we assumed here.  Figure \ref{fig:rcore} compares the evolution
of the core radii in simulations \#2, \#3 and \#4 with the core radii
of the clusters in the Mackay \& Gilmore sample.  The core radii in
the LMC sample are determined by fitting the observed intensity
profiles with 2-parameter King (1966) models, the parameters being
half-mass radius and concentration.  The core radius in these models
is generally about a factor of four larger than that defined by
theorists.\footnote{The observer's core radius is traditionally where
the surface brightness drops by a factor of two.  The ``continuum''
theorist's core radius is defined in terms of the central density and
velocity dispersion, not from the density variation, although the
falloff is implicit in the solution to Poisson's equation.  The N-body
theorist's core radius is something different again, defined in terms
of local density from the criterion proposed by
\cite{1985ApJ...298...80C}.  Remarkably, they all seem to have
something to do with one another.}  We correct for this discrepancy by
dividing the core radii provided by Mackay \& Gilmore by a factor of
four.

Except for the star cluster R136 (the central cluster of NGC 2070;
leftmost point) the observed core radii appear to be quite consistent
with the core evolution of simulations \#2--4. The core radius of
R\,136 is consistent with that of simulation \#1 (partially shown
in Figure\,\ref{fig:rcore}) .

\subsection{Evolution of the mass function}\label{Sect:mass_function}

All simulations started with a Salpeter (power-law with exponent
$\alpha = -2.35$) mass function. The global mass function changes with
time due to stellar evolution and selective evaporation driven by the
dynamical evolution of the cluster.  The mass function also varies
locally due to mass segregation.  To quantify the global and local
changes in the stellar mass function, we first select those stars
which remain on the main sequence during the entire period studied.
The turn-off mass at 100\,Myr in our stellar evolution model is about
4.6\,\msun.\nocite{1989ApJ...347..998E} By restricting ourselves to
main-sequence stars in the rather narrow mass range of 1 to
4.6\,\msun\, we guarantee that the mass function is not affected by
blue stragglers, giants or stellar remnants, although some
contamination from dormant blue stragglers cannot be avoided (see
\S\,\ref{Sect:BlueStragglers}).

\begin{figure}
a)  \psfig{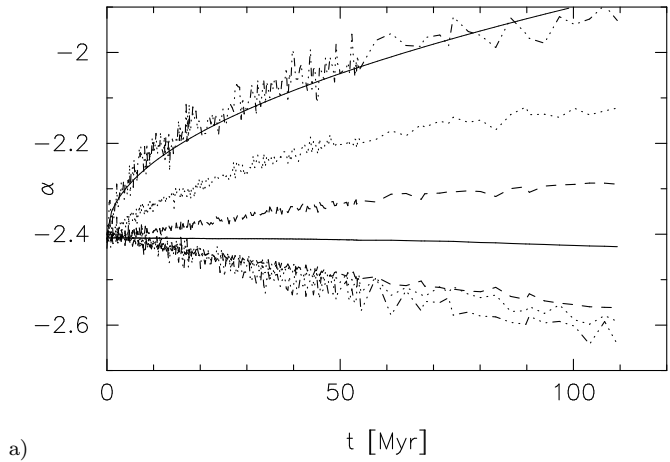}
   \caption[]{Evolution of the power-law exponent $\alpha$ of the mass
   function for simulation \#1 for main-sequence stars of masses
   between 1 and 4.6\,\msun. The solid almost horizontal curve
   represents the entire cluster.  The curves above it (flatter mass
   function) give the value of the power-law slope to the mass
   function of the stars which are {\em inside} the 50\% (dashes),
   25\% (dotted line) and 10\% (dash-3dotted line) Lagrangian radii.
   The curves below the solid curve in the middle (steeper mass
   function) give the value of the power-law slope to the mass
   function of the stars which are {\em outside} the 50\%, 25\% and
   10\% Lagrangian radii.  The thin solid curve through the upper
   dash-3-dotted curve (mass function of the stars within the 10\%
   Lagrangian radius) is calculated using $\alpha \propto \sqrt{t}$
   (see text).  Similar expressions fit the lower curves. (Note that
   after 55\,Myr we increased the output time interval from 1 Myr to 5
   Myr). } \label{fig:Cimffit}
\end{figure}

Figure\,\ref{fig:Cimffit} shows the evolution of the cluster mass
function exponent between 1 and 4.6\,\msun\, for simulation \#1. The
exponent was calculated using a least-squares fit to the binned (100
bins) mass data.  From top to bottom, the curves present the mass
function exponent (i) within the 10\% Lagrangian radius (together with
an analytic fit, as described below), (ii) within the 25\% Lagrangian
radius, (iii) within the half-mass radius (50\% Lagrangian radius),
(iv) for the entire cluster, (v) outside the half-mass radius, (vi)
outside the 75\% Lagrangian radius, and (vii) outside the 90\%
Lagrangian radius.  The global mass function (lower solid curve)
steepens slightly with time, from $\alpha \simeq -2.41$ at birth to
about $\alpha \simeq -2.43$ at an age of 100\,Myr.  Note that the
value of the mass function exponent is not -2.35 because of the
presence of binaries. The mass function for the inner 10\% of the
stars (upper dash-3-dotted curve in Figure\,\ref{fig:Cimffit}) is most
strongly affected by mass segregation.

The slight steepening of the global main-sequence mass function is the
result of dynamical activity in the cluster center, which tends to
eject relatively high mass stars from the cluster more frequently than
lower mass stars, simply because the latter are not as abundant in the
cluster core, and hence do not participate so frequently in strong
dynamical encounters. Thus the change in the global mass function is
driven mainly by dynamical encounters in the cluster core, and not by
selective evaporation (but see \cite*{2006A&A...452..131L}). This is
in part because of our neglect of a global tidal field.  At about
60\,Myr the global mass function stops changing because the white
dwarfs formed subsequently are comparable in mass to the least massive
main-sequence stars in the simulation, and compete with them
dynamically.

The mass functions for the stars in the outer parts of the cluster
become steeper with time, with power-law exponents ranging from
$\alpha \simeq-2.56$ for the outermost 50\% to $\alpha \simeq-2.64$
for the outer 10\% Lagrangian radius in simulation \#1.  The mass
functions in the inner parts of the cluster become flatter with time.
Within the half-mass radius the mass function flattens to $\alpha
\simeq -2.3$ in about 100\,Myr; for the inner 10\% Lagangian radius
the mass function flattens to $\alpha \simeq -1.9$. The other simulations
had considerably smaller changes in the global mass function.

We fit the variation in the mass function exponent in the inner part of
the cluster by
\begin{equation}
    \alpha(t) = \alpha(0) + \left(t \over \tau\right)^{0.5}
\end{equation}
\citep{2004ApJ...608L..25M}.  Here $\alpha(0)$ is the initial
power-law slope and $\tau$ is a constant.  For stars within the inner
10\%, 25\%, and 50\% of the simulation (top three dashed/dotted curves
in Figure \ref{fig:Cimffit}), the best-fitting values of $\tau$ are
$\sim 5\trlx, 15\trlx$, and $130\trlx$, respectively.  The fit for
the inner 10\% Lagrangian radius is presented in
Figure\,\ref{fig:Cimffit} as the upper thin solid curve. The mass
functions in the other runs vary similarly, with the same scaling
($\propto \trlx$) but larger coefficients. 


\subsection{Evolution of the stellar population}

\begin{figure}
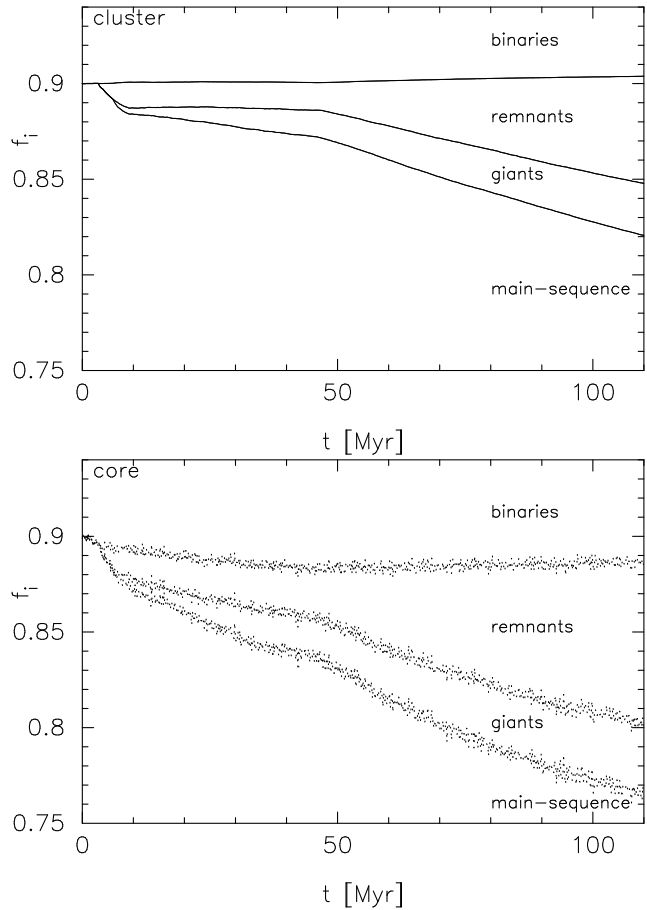

   \psfig{figure=./fig_B_FracStarvsT2b.ps,width=\columnwidth,angle=-90}
   \psfig{figure=./fig_B_FracStarvsT2a.ps,width=\columnwidth,angle=-90}
   \caption[]{Stellar content of simulation \#2 as a function of time.
   We distinguish here between main-sequence stars, giants, remnants
   (single white dwarfs, neutron stars and black holes) and binaries.
   The regions between the various lines indicate the fraction of each
   category of object, as is indicated for the solid curves to the
   right of the figure.  The {\bf top panel} give these branching
   ratios for the entire cluster (in solid curves), while the dotted
   lines in the {\bf bottom panel} show the branging ratios for the
   inner 10\% Lagrangian radius. The various stellar groups are
   represented in the same order, but due to the different evolution
   in the cluster interior they are shifted with respect to the solid
   curves.
   \label{fig:B_Stellarcontent} }
\end{figure}

As the cluster ages, main sequence stars turn into giants, which then
evolve into stellar remnants.  This is illustrated in
Figure\,\ref{fig:B_Stellarcontent}, which shows the stellar content of
simulation \#2.  Here we distinguish between binaries (top part of the
diagrams) and single stars (bottom part).  For the single stars, we
make a further division into main sequence stars, giants and compact
objects. The total is normalized to unity so that the diagram gives
the relative fraction of each type of object as a function of time.
The solid curves (top panel) represent the population of the entire
star cluster; dotted curves (bottom panel) present the same population
information for the innermost 10\% of the system.

Initially, the cluster contains binaries and main-sequence stars, but
the composition changes as stars evolve, and by an age of about
10\,Myr a significant population of giants and stellar remnants is
established. Between 10 and $\sim 50$\,Myr the fraction of giants
gradually increases at the cost of the main-sequence stars, while the
fraction of stellar remnants hardly changes.  The main reason for the
constancy of the fraction of stellar remnants is that the neutron
stars, which are born during this time interval, tend be ejected from
the cluster upon formation (see \S\,\ref{Sect:CompactObjects} below).
After about 50\,Myr the relative fractions of giants and remnants
increase more rapidly than before, primarily because of the formation
of white dwarfs (see also Figure\,\ref{fig:B_Nbin_rmn}).

The relative fraction of binaries in the inner parts of the cluster
(top curve in the bottom frame of Figure\,\ref{fig:B_Stellarcontent})
is larger than for the cluster overall, and the same is true for
remnants and giants. The reason for this is that these objects are all
more massive than the average main-sequence star, and therefore sink
to the center by mass segregation.

One of the most interesting features in Figure
\ref{fig:B_Stellarcontent} is the roughly constant proportion of
binaries. The total binary fraction drops by only about 10\% percent
during the 100\,Myr of the evolution.  For the other models, the
binary fraction falls from its initial value of 10\% to 9.1\% for
model \#2, 8.9\% for model \#3 and to 9.0\% for model \#4.  This
result is consistent with the findings of \cite{2006ApJ...646..464S},
who observed only a slight decay in the initial binary fraction in
their simulations of the the cores of dense star clusters.

\subsection{Characteristics of the blue straggler population}
\label{Sect:BlueStragglers}

All the stars in our simulated clusters are born on the zero-age
main-sequence.  The normal evolution of a star may be altered by mass
transfer from a more massive Roche-lobe filling companion, or by
collisions with other stars.  Mass gained from a companion causes an
accretor to become more massive, and may also refresh the nuclear fuel
in the stellar interior by mixing central helium with fresh hydrogen.
Internal mixing tends to rewind the nuclear clock of the star.  A
physical collision between two main sequence stars may have a similar
effect
\citep{1997ApJ...484L..51S,2001ApJ...548..323S,2002ApJ...568..939L,2005MNRAS.358..716S}.
Either effect may ``rewind'' the star's nuclear clock, and such
rejuvenated stars can appear on a Hertszprung-Russell diagram as blue
stragglers.  For definiteness, we identify a blue straggler as a
main-sequence star with mass larger than the turn-off mass at the
current cluster age.

\begin{figure}
   \psfig{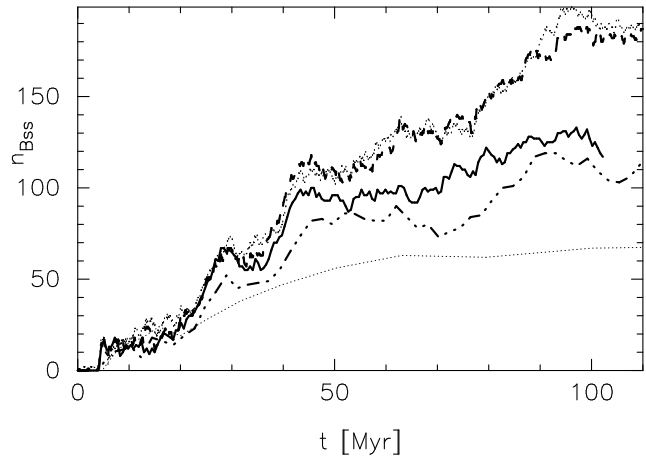}
\caption[]{Evolution of the number of blue stragglers in simulations
\#1\, (dotted curve), \#2\, (dashes), \#3\, (solid curve), \#4\,
(dash-3-dotted curve). The results of the binary population synthesis
of model A17 by \cite{1994A&A...288..475P} are presented as the lower
thin dotted line.
\label{fig:NBss_ABCDvsT}
}
\end{figure}

In each simulation an appreciable number of blue stragglers formed,
mostly through mass transfer in close binaries.  Collisions between
main-sequence stars are rare, except in simulations \#1R and \#1,
where they give rise to a collision runaway.  Run \#1R in its first
12\,Myr is identical to one of the simulations described in
\cite{2004Natur.428..724P}.  In such dense systems, repeated
collisions can result in an unusually massive blue straggler
\citep{2001MNRAS.323..630H,2003MNRAS.345..762L} within the first few
million years.  A collision between a main sequence star and an
evolved (subgiant) star generally does not result in the formation of
a blue straggler; a bright giant is a more common outcome
\citep{1997A&A...328..143P}.

Figure\,\ref{fig:NBss_ABCDvsT} presents the evolution of the total
numbers of blue stragglers in simulations \#1 through \#4.  The number
of blue stragglers in each simulation increases steadily with time.
Up to an age of about 50\,Myr, the number increases at a similar rate
in each of the simulations.  After $\sim 50$\,Myr, the number in the
two shallower clusters (\#3 and \#4) continues to increase, but at a
lower rate. The growth rate of the number of blue stragglers in these
models is consistent with expectations based on the binary population
synthesis calculations of \cite{1994A&A...288..475P}.

In the population synthesis study of \cite{1994A&A...288..475P},
gravitational interactions with other cluster stars were ignored.  In
their model A17 they adopted a similar mass function to ours, but with
a minimum mass of 1.47\,\msun\, instead of our minimum mass of
1.0\,\msun\, (see \S\,\ref{Sect:Simulation}), and they included a
slightly wider binaries than in our simulation \#4.  They simulated
17000 objects with a 75\% binary fraction, resulting in 12\,750
binaries, comparable to the 13\,107 binaries in our models \#1 to \#4.
The differences in initial conditions are expected to increase the
number of blue stragglers in our simulation \#4 by at most a factor of
two compared to model A17 of \cite{1994A&A...288..475P}.  In
Figure\,\ref{fig:NBss_ABCDvsT} we reproduce the evolution of the
number of blue stragglers in their model A17.  The results of our
simulation \#4 (dash-3-dotted curve) is consistent with their model
A17.  We continued simulation \#4 to an age of about 400\,Myr to
confirm that also at later time ($t>100$\,Myr) the number of blue
stragglers in this simulation is consistent with the results of
simulation A17.

\begin{figure}
   \psfig{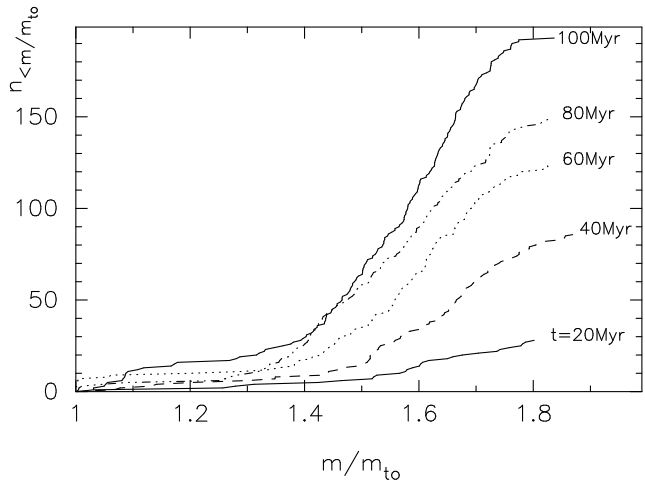}
\caption{Cumulative distribution of masses above the turn-off mass for
blue stragglers at various instants during simulation \#1.  The mass
at the turn off is $m_{\rm to} \simeq 10.76$\,\msun\, at an age of
20\,Myr (lower solid line in Figure\,\ref{fig:NBss_A}), $m_{\rm to}
\simeq 7.29$\,\msun\, at 40\,Myr (dashes), 6.03\,\msun\, at 60\,Myr
(dots), 5.15\,\msun\, at 80\,Myr (dash-3dots) and 4.63\,\msun\, at
100\,Myr (top solid curve).
\label{fig:NBss_A}
}
\end{figure}

To allow comparison of the characteristics of the blue stragglers at
different times we consider the quantity $m/m_{\rm to}$, the mass of a
blue straggler normalized to the instantaneous turn-off mass.
Figure\,\ref{fig:NBss_A} presents the cumulative distribution of
$m/m_{\rm to}$ at various moments in time during the evolution of the
densest cluster, simulation \#1.  In this simulation, only $\aplt
20$\% of the blue stragglers have masses less than 40\% above the
turn-off; most have $m/m_{to}\sim1.4-1.7$.  The main difference
between these curves is simply the numbers of blue stragglers---the
shapes of the normalized distributions are almost indistinguishable.

\begin{figure}
   \psfig{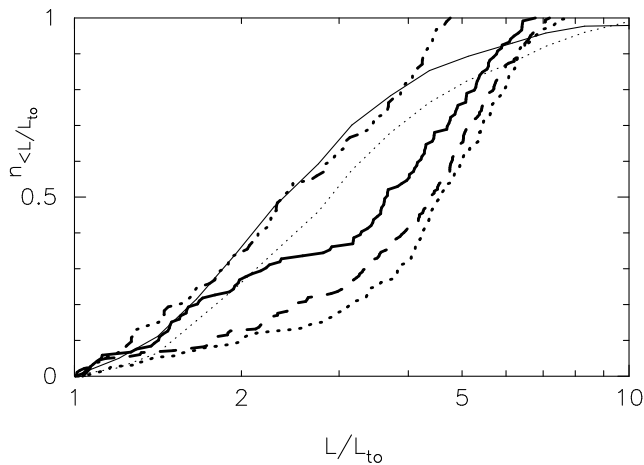}
\caption{Cumulative distribution of $L/L_{\rm to}$ for simulations
\#1\, (dotted curve), \#2\, (dashes), \#3\, (solid curve) and \#4\,
(dash-3-dots).  The thin solid and dotted curves show the cumulative
distributions of the blue stragglers and bright ($M_v<-8.8$) and faint
($M_v>-8.8$) globular clusters, from the data of Piotto et al (2004).
The thin solid line appears to be consistent with the thick
dash-3dotted curve, but bear in mind here that one is for a 100\,Myr
old population, whereas the other is for a 10\,Gyr old population.
\label{fig:LBss_ABSD}
}
\end{figure}

In Figure\,\ref{fig:LBss_ABSD} we show the cumulative distribution, at
an age of 100 Myr, of $L/L_{\rm to}$, the blue straggler luminosity
relative to the luminosity of a main-sequence star at the turn-off.
These distributions change only slightly with time.  Most striking in
Figure\,\ref{fig:LBss_ABSD} are the run-to-run variations: a decrease
in cluster density causes the median blue straggler mass to drop
gradually from $m/m_{\rm to} \simeq 1.6$ for simulation \#1\, to
$m/m_{\rm to} \simeq 1.4$ for simulation \#4.  As dynamical effects
become more important (progressing from simulation \#4 to \#3, \#2 and
finally \#1), the blue straggler luminosity distribution shifts to
higher luminosities.  This is a direct consequence of the changing
importance of the main formation channels: binary evolution and
stellar collisions.  The latter channel is unimportant in low-density
clusters, but dominates in the densest systems.

In our most compact model cluster, simulation \#1, the majority of
blue stragglers are the result of dynamical interactions---binary
hardening and exchange---and tend to be more massive than blue
stragglers formed by binary evolution.  Binary hardening boosts the
number of blue stragglers by shifting the moment of Roche-lobe contact
to an earlier evolutionary stage of the primary.  In closer binaries,
more of the donor mass can be transferred to the accretor without
spillage, whereas in wider binaries most mass is lost from the binary
system.  Exchange interactions tend to substitute the lower mass
binary component for a higher mass incoming star, with the result that
post-encounter binary components tend to be more similar in mass than
in the pre-encounter system.  The binaries are generally also hardened
in the process.  Mass transfer in a binary consisting of two stars of
comparable mass proceeds more conservatively than in unequal mass
binaries.  Thus, the higher encounter rate in denser systems has two
distinct effects: it boosts the formation rate of blue stragglers, and
also leads to more massive blue stragglers.  However, we should bear
in mind that, with $\aplt 200$ blue straggler present at any time
during the first 100\,Myr in our simulation models, their formation
remains a rare event even in very dense clusters.

Overplotted in Figure\,\ref{fig:LBss_ABSD} are two cumulative
luminosity profiles from the observed sample of 2798 blue stragglers
in 56 Galactic globular clusters from \cite{2004ApJ...604L.109P}.
They divided the clusters in which they found blue stragglers into two
groups, one for ``faint'' clusters ($M_v > -8.8$; thin solid curve),
the other for ``bright'' clusters ($M_v<-8.8$; thin dotted curve).
The blue stragglers found in the denser clusters are typically
brighter than those in the less dense clusters
\cite{2004ApJ...604L.109P}.  The brightness distribution of the blue
stragglers present at an age of 100\,Myr in our simulation \#4 is
strikingly similar to the distribution of the observed blue stragglers
in low density clusters, even though the ages are completely different
(100\,Myr for simulation \#4, compared to about 10\,Gyr for the
globular cluster population).  The main difference is at the
high-luminosity end, which is in part a consequence of the larger
relative brightness of the lower mass stars in the sample used by
\cite{2004ApJ...604L.109P}: at an age of 100\,Myr, a star twice as
massive as the turn-off is about an order of magnitude brighter than
the turn-off luminosity, whereas at an age of 10\,Gyr the luminosity
difference is about a factor of 20.

In our simulation models \#1--3\, the number of collisions is
probably much higher than in the old globular clusters used to compile
the list of blue stragglers included in the thin dotted curve.  The
trend found by \cite{2004ApJ...604L.109P} that the blue stragglers
formed in the more massive clusters are typically brighter is also
present in our simulations.  We find that clusters with a higher
stellar density produce brighter blue stragglers. This is consistent
with the expectation of \cite{2004MNRAS.349..129D}.  In that case, we
confirm that the higher encounter rate in the denser stellar
environments contributes to the brighter blue stragglers.

\begin{figure}
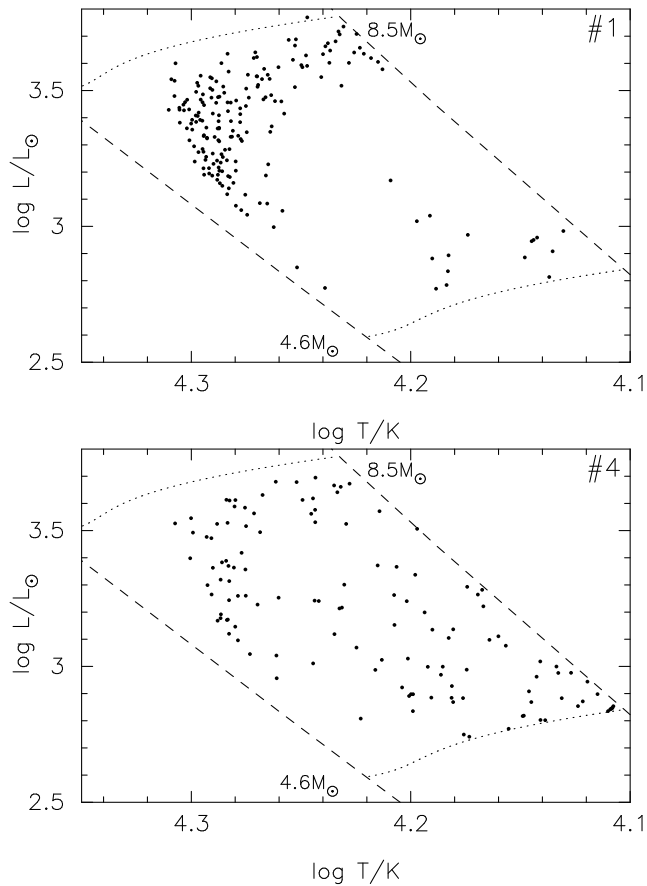

   \psfig{figure=./fig_A_TLBss.ps,width=\columnwidth,angle=-90}
   \psfig{figure=./fig_D_TLBss.ps,width=\columnwidth,angle=-90}
   \caption[]{Temperatures and luminosities of the blue stragglers in
   simulations \#1\, and \#4\, at an age of 100\,Myr.  The two
   vertical dashed lines represent the zero-age main-sequence and the
   terminal age main-sequence. The two dotted curves are the
   main-sequence tracks for stars of 4.6\,\msun\, (lower curve) and
   8.5\,\,\msun\, (upper curve). 
   \label{fig:AD_TLBss} }
\end{figure}

Blue stragglers found in denser clusters are generally bluer than
those in low density clusters \citep{2004ApJ...604L.109P}.  This
effect is also evident in Figure\,\ref{fig:AD_TLBss}, which compares
the temperatures and luminosities of the blue stragglers in simulations
\#1 and \#4 at an age of 100\,Myr.  The denser model clusters tend to
produce more massive and therefore bluer blue stragglers.  We note
that, even though all blue stragglers are (by definition)
main-sequence stars, they have a broad range of temperatures for any
given luminosity, because of the spread in their
effective ages. The blue
stragglers formed in the densest clusters have the broadest range in
ages, and are almost uniformly distributed between the zero-age and
terminal-age main sequence.

\subsection{Future blue stragglers}

During the early evolution of our simulated clusters, a relatively
large number of low-mass ($m \ll m_{\rm to}$) stars receive mass from
a companion star or participate in a merger.  The resultant star will
be rejuvenated, but so long as its mass lies below the turn-off it
will not be identifiable as a blue straggler.  Low-mass stars which
accrete mass at an early stage appear as blue stragglers once the
cluster has aged sufficiently to allow them to lag behind the main
sequence turn-off.  The early dynamical evolution of the cluster can
thus lead to a reservoir of potential future blue stragglers, which
become evident only later in the cluster's evolution.  In each
simulation, the number of such potential blue stragglers
(specifically, rejuvenated main-sequence stars with $m < 4.6$\,\msun,
which corresponds to the turn-off mass at about 100\,Myr, see
\S\,\ref{Sect:mass_function}) was $250\pm20$, where the error
indicates the run-to-run variation. (We include here the additional
constraint that a star must spend at least $10^5$ years above the
turn-off to be included.)  The blue straggler mass distributions are
indistinguishable from one run to another: they are consistent with a
constant number of potential blue stragglers per unit mass.

This lack of a trend in the properties of potential blue stragglers
with cluster parameters is opposite to the trends seen in the actual
blue stragglers identified during the first 100\,Myr, where a clear
dependence on cluster density is evident (see
Figure\,\ref{fig:LBss_ABSD}).  The reason is that most potential blue
stragglers stem from evolving primordial binaries with relatively
low-mass secondaries.  Such binaries are not likely to be dynamically
active, and their orbital parameters are largely unaffected by the
dynamical evolution of the parent cluster.  In our simulations, the
secondary stars in binaries were initially distributed uniformly in
mass between the minimum mass and the mass of the (pre-selected)
primary (see \S\,\ref{Sect:Simulation}).  The mass distribution of
potential blue stragglers therefore simply follows the initial
distribution of secondary masses.

The once dormant blue stragglers generally spend relatively little
time as main-sequence stars above the turn off.  Half ($\sim 120$)
live for less than 10\,Myr, and fewer than 10\% live between 30\,Myr
and 100\,Myr as blue stragglers.  Interestingly, the more massive
stars ($\apgt 2.5$\,\msun) tend to spend most time as blue stragglers.
This can also be understood from binary evolution.  Dormant blue
stragglers in our runs are formed from mass transfer in primordial
binaries of which the primary typically has $m > 4.6$\,\msun.  It is
typically the secondary which eventually turns up as a blue straggler,
and binaries with comparable component masses tend to produce the most
massive blue stragglers.  A very low-mass secondary will generally
accrete little mass in a phase of mass transfer, resulting in a blue
straggler that spends only a short time above the turn-off, and never
very far above it.  A star is rejuvenated more effectively when it has
accreted more mass and when it is evolved further along its main
sequence, both effects resulting in a longer blue straggler lifetime.

\subsection{OB runaway stars}

The Galaxy contains a population of stars having velocities
$>40$\,km\,s$^{-1}$.  These objects are generally called OB runaways
\citep{1954ApJ...119..625B,1993msli.conf..207B}, because of their high
velocities and predominant spectral type.  There are two main theories
to explain these high velocity objects: ejection from a binary system
as the companion experiences a supernova, and dynamical ejection
following a close multi-body encounter \citep{1995MNRAS.277.1080L}.
The first mechanism was studied extensively by means of binary
evolution of individual cases \citep{1996A&A...305..825V} and by
synthesizing entire Galactic binary populations
\citep{2000ApJ...544..437P}.  The dynamical ejection mechanism has
attracted considerable attention in the past few years, and has been
studied for individual cases \citep{2004MNRAS.350..615G} and in
population studies \citep{1995MNRAS.277.1080L}.  Neither theory
satisfactorily explains the origin of most runaways.

 From an observational point of view, both scenarios seem to operate
concurrently.  A large number of supernova-produced runaways were
identified by \cite{2000astro.ph.10057H} using the Hipparcos database,
and individual cases were also revealed to originate from this
scenario \citep{1997ApJ...475L..37K}, but some clear cases must have
their origin in dynamical ejection \citep{1986ApJS...61..419G}.
Clearly, the jury is still out on the mechanism responsible for OB
runaway stars.  The majority of runaways are high-mass stars.
Furthermore, some 40\% of all stars of spectral type O are runaways,
and 25\% of all B stars, but only 4\% of A stars
\citep{1954ApJ...119..625B,1993msli.conf..207B}.

In our simulations we encountered quite a number of ejected stars
having velocities high enough for them to be considered runaways.
Within 100\,Myr, a total of $6950\pm150$ stars were ejected from each
cluster.  The number of stars ejected at high speed does not seem to
depend on the initial cluster density.  The denser clusters, however,
do tend to eject more main-sequence stars and at an earlier age, while
less concentrated clusters tend to eject more white dwarfs.  In
simulation \#1, a total of 815 main-sequence stars were ejected with
high ($>40$\,\kms) velocity, compared to 290 from simulation \#2, 103
from \#3 and 67 from simulation \#4.  The number of ejected high-speed
binaries tends to increase with cluster density, from about 150 for
the shallowest cluster (\#4) to about 200 for the most concentrated
model (\#1).  Only 2--3\% of the ejected stars are members of
binaries, whereas the overall binary fraction initially is 10\%.

At zero age, our models contained 2695 stars of spectral type O9.5 or
earlier. With a total of 444 runaways in simulation \#1, including
less than 100 O stars, we find fewer than 4\% runaways among the
spectral type O stars. Thus it seems that even in the densest clusters
OB runaways are underproduced by about an order of magnitude compared
to observations.  Note in addition that the runaways on average spend
about 60--80\% of their main-sequence lifetimes in the cluster before
acquiring high velocities, so the discrepancy with the observations is
actually even larger.  Lower mass stars tend to spend more time in the
cluster before they are ejected.  The fraction of binaries among the
OB runaways is 12\%, considerably higher than the overall binary
fraction among ejected stars, and comparable to the initial binary
fraction.

\begin{figure}
   \psfig{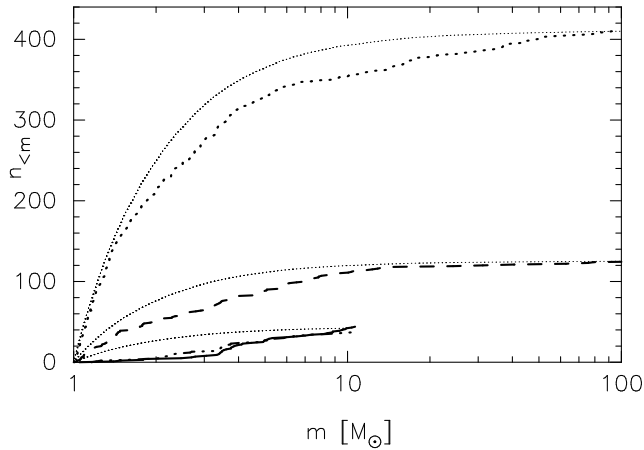}
   \caption[]{Cumulative mass distribution of runaway stars ($v>40$
   km\,s$^{-1}$) for simulations \#1 (dotted curve, 444 stars), \#2,
   (dashes, 131 stars) and (combined) models \#3 and \#4
   (dash-3-dotted curve, 46 and 38 stars, respectively).  The Salpeter
   initial mass function used as initial conditions for the
   simulations is presented as the thin solid curve.  Only stars which
   escape within 100\,Myr with velocity exceeding 40\,km/s are shown.
   \label{fig:MmsVesc} }
\end{figure}

Figure\,\ref{fig:MmsVesc} shows the cumulative mass distribution of
stars which escaped our simulated clusters with velocities exceeding
$40$\,km\,s$^{-1}$.  For comparison we also plot the adopted Salpeter
initial mass function (thin solid curve).  Evidently, the mass
distribution of runaway stars from the most concentrated clusters tends
to follow the initial mass function in these models, while shallower
models contain a much greater proportion of high-mass stars.

The ejection of compact objects was discussed in
\S\,\ref{Sect:CompactObjects}, but in our simulations a sizable number
of massive stars are ejected before exploding as supernovae. These
high-velocity stars will therefore produce supernovae in the
interstellar medium.  For simulated clusters \#1, \#2, \#3 and \#4,
respectively, a total of 78, 26, 10 and 4 stars with $m >8$\,\msun\,
were ejected before becoming supernovae.  In simulation \#1, half (40)
of these travel a distance of $\sim0.5$\,kpc before exploding; the
maximum distance was 4.4kpc.

\subsection{Evolution of compact objects}\label{Sect:CompactObjects}
\label{Sect:retention}

Black holes and neutron stars receive kicks upon formation in our
simulations.  These natal kicks can be quite high
\citep{1987ApJ...321..780D,1994Natur.369..127L,2006ApJ...639.1007W}
and we select them (with random direction) from a Paczynski-Hartman
distribution \citep{1990ApJ...348..485P,1997A&A...322..127H}, which
has a dispersion of $\sigma_{\rm kick} = 300$\,\kms.  Black holes seem
to receive on average lower kicks
\citep{1996ApJ...473L..25W,2005ApJ...618..845G}, and in our
simulations a black hole of mass $m_{\rm bh}$ receives a kick drawn
from the same distribution as the neutron star kicks, but reduced by a
factor of $1.4\,\msun/m_{\rm bh}$.

Table\,\ref{Tab:compactobjects} presents the number of core collapse
supernovae and compact objects produced for the simulations listed in
Table\,\ref{Tab:initials}.  In each run, about 1700 black holes form
in Type Ic supernovae, and 6500 neutron stars form in Type Ib or Type
II supernovae.  Although the various models share the same realization
of the initial mass function, the numbers of supernovae differ
slightly from run to run; with more supernovae occurring in the
larger, less dense clusters. This is caused mainly by variations in
binary evolution, due to the differences in binary separations and
cluster density between runs.  Also, for individual supernovae the
kick is applied randomly, with different random seeds for different
runs.  This introduces an unbiased variation in the evolution of the
massive binary population, giving rise to additional slight
differences in the numbers of compact objects that remain bound to
their parent cluster.

The lower kicks imparted to black holes mean that a smaller fraction
of them escape the cluster.  The retention fraction for black holes
ranges from 0.48 to 0.71 (see Table\,\ref{Tab:compactobjects}); for
neutron stars, the range is 0.065 to 0.12.  The densest clusters
retain most compact objects, because of their larger escape speeds.
The range of retention fraction is about a factor of two, while the
densest cluster has an escape velocity roughly four times greater than
that of the least dense cluster.

Most compact objects are single, since supernovae are very effective
at destroying binaries.  At an age of 100 Myr, only about 4\% of black
holes and $\sim 2$\% of neutron stars are members of binary systems.
These fractions are lower than the overall binary fraction $\sim10\%$)
mainly due to evolution of the binary companion, and also because of
differences in the amount of mass lost during the supernova---the
formation of a neutron star is generally associated with considerably
more mass loss than is the creation of a black hole.  The companion of
a black hole is generally rather massive and may also explode in a
supernova to form another black hole or a neutron star, but only a few
binaries survive both supernovae.  No binary containing two neutron
stars was formed in any of our simulations, whereas black holes tend
to be paired with other black holes (see
Table\,\ref{Tab:compactobjects}).

Black holes are also more likely than neutron stars to have a main
sequence star or a giant as a companion.  Most of these binaries,
indicated by (bh,ms) in Table\,\ref{Tab:compactobjects}, will become
X-ray binaries at some point during their evolution.  Among these
binaries, black-hole accretors are about twice as common as
neutron-star accretors. We therefore expect that in star clusters with
ages $\sim 100$\,Myr, X-ray binaries containing a black hole may be
quite common, possibly even more common than X-ray binaries with a
neutron star.

\begin{table*}
\caption[]{Some characteristics of the compact object population at
100Myr. The first column identifies the run, followed by the total
number of core collapse supernovae that result a black hole (Ic) or a
neutron star (Ib+II). The next two columns give the number of black
holes and neutron stars still present in the cluster at an age of
100\,Myr. The following columns give the number of binaries containing
two black holes (bh,bh), a black hole and a neutron star (bh,ns), or a
white dwarf (bh,wd) and (ns,wd).  The final two columns give the
number of black holes and neutron stars that are accompanied by a
main-sequence star; note that the few that are accompanied by an
evolved giant star are also included in this category.}
\bigskip
\begin{tabular}{lrrr|rrrrrrr}
Simulation   & Ic  &Ib+II& $N_{bh}$ & $N_{ns}$ &(bh,bh) &(bh,ns) 
	& (bh, wd) & (ns, wd)
                     & (bh, ms)& (ns, ms) \\
\hline		   			 	      			    
\#1R   & 1632 & 6520 & 1152 & 811 & 14 & 8 & 1& 7& 10 & 13 \\
\#1    & 1656 & 6584 & 1017 & 585 & 12 & 1 & 1& 7& 13 &  5 \\
\#2    & 1710 & 6446 & 1028 & 553 & 16 & 4 & 7& 5& 10 &  1 \\
\#3    & 1717 & 6387 &  834 & 365 & 17 & 2 & 9& 4&  9 &  3 \\
\#4    & 1728 & 6735 &  828 & 436 & 11 & 4 & 5& 9& 13 &  5 \\
\label{Tab:compactobjects}
\end{tabular}
\end{table*}


\begin{figure}
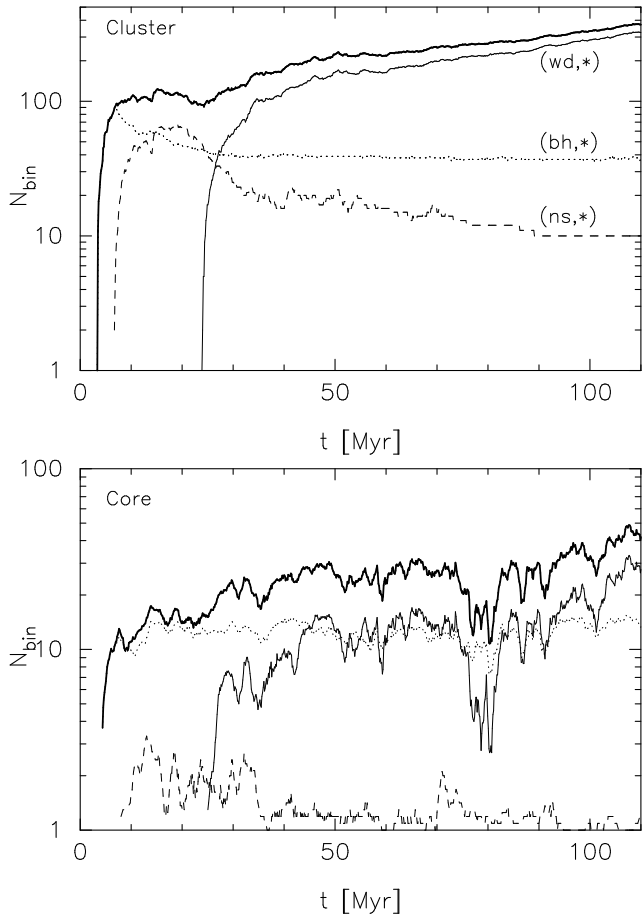

   \psfig{figure=./fig_B_Nbin_rmn.ps,width=\columnwidth,angle=-90}
   \psfig{figure=./fig_B_Nbin_core_rmn.ps,width=\columnwidth,angle=-90}
   \caption[]{Number of binaries containing a compact object as a
   function of time for simulation \#2.  The upper panel presents data
   for the entire cluster, whereas the lower panel includes only
   binaries within $2\rcore$.  In each panel, the top (thick solid)
   curve gives the total number of binaries containing at least one
   compact object (bh, ns or wd). The thin solid curve gives the
   number of binaries containing one white dwarf, dashes are for
   neutron stars, and dots are for binaries containing at least one
   black hole.  Note that due to double counting the total of the thin
   curves does not add up to the thick curve.  \label{fig:B_Nbin_rmn}
   }
\end{figure}

Figure\,\ref{fig:B_Nbin_rmn} presents, for simulation \#2, the numbers
of binaries containing at least one compact object, as functions of
time.  The number of black holes in binaries rises sharply shortly
after the start of the simulation, with a peak at around 8\,Myr.  This
is the moment when the turn-off mass drops below the minimum mass for
forming black holes ($\sim 23$\,\msun), and lower mass stars form
neutron stars.  The number of black holes in binaries drops rapidly
from this moment on because many of their companions form neutron
stars in supernova explosions (see Table\,\ref{Tab:compactobjects}).
This transition is also visible in the sharp increase at 8 Myr of the
number of binaries containing a neutron star.  Note that binaries
containing a neutron star and a black hole are counted twice in this
figure, as both (bh, $\star$) and (ns, $\star$).

White dwarfs become significant components of the compact binary
population after about 25\,Myr, at a turn-off mass of about
10\,\msun. Stars of $\aplt 8$\,{\msun} evolving in isolation turn into
white dwarfs, but in a binary system early stripping of the hydrogen
envelope may cause a more massive star to become a white dwarf instead
of collapsing to a neutron star. The population of compact binaries in
clusters older than about 40\,Myr is dominated by white dwarfs. The
naked cores of stars which are stripped of their hydrogen envelope,
exposing their hot helium interior, are also included in this
category, but Wolf-Rayet stars are not.  The reverse process may also
occur: If a sufficient amount of mass is accreted by (say) a
7\,\msun\, star it may, by the end of its fuel processing life, still
collapse to a neutron star in a supernova explosion, whereas under
normal circumstances it would become a white dwarf.  A similar process
may result in the formation of a black hole from a star of mass $\aplt
23$\,\msun.

One interesting evolutionary product of such a process is a binary in
which a neutron star is accompanied by a white dwarf in an eccentric
orbit.  Such a population, predicted by binary population synthesis
models \citep{1999MNRAS.309...26P}, seems to be quite common in our
simulations.  Although the statistics are poor, approximately half (18
of 32) of our (ns, wd) binaries have quite high eccentricities: $e =
0.45 \pm 0.27$.  In only one of these cases (in simulation \#1) was
the eccentricity induced by a strong dynamical encounter with another
cluster member; in all others it was the result of reverse evolution,
as described by \cite{1999MNRAS.309...26P}: For sufficiently massive
binaries (with a $\apgt 8$\,\msun\, secondary), mass transfer can
result in a reversal of the order in which the component stars become
supernova \citep{1994A&A...290..119P}, causing the less massive
(secondary) star to explode before the primary.  If the mass of the
secondary star is just below the limit for forming a black hole, such
a reversal in the supernova order may lead to binaries with a mildly
recycled millisecond pulsar orbiting a black hole
\citep{2004MNRAS.354L..49S}. In our simulations we find a total of 5
such cases out of 19 binaries in which a black hole is accompanied by
a normal radio pulsar, much larger than predicted from the simulations
of \citep{2004MNRAS.354L..49S}.  The difference in formation rate
stems in large part from differences in initial conditions.
The maximum initial separation in our simulations was chosen based on
the hard-soft boundary \citep{1975MNRAS.173..729H} for that particular
cluster, whereas Sipior et al.\, adopted a maximum separation based on
observed binaries in the solar neighborhood.

\subsection{Gravitational wave sources and type Ia supernovae}

The compact object binaries in our simulated clusters may at some time
become sources of gravitational waves.  The binaries containing back
holes or neutron stars are not very promising gravitational wave
sources because they are rare (see however
\cite*{2006astro.ph..3441K}), but a sizable fraction of binaries with
two white dwarfs may become important sources for space-based
gravitational wave detectors
\citep{2004MNRAS.349..181N,2005ApJ...633L..33S}.  In order to improve
the statistics we combine the white dwarf binaries formed in
simulations \#1R, \#1 and \#2, and those in simulations \#3 and \#4.
The total number of white dwarf binaries that engage in Roche-lobe
overflow within 10\,Gyr is $21\pm2$ per cluster in the denser clusters
(\#1 and \#2), and $31\pm1$ in the shallower clusters (\#3 and \#4).


\begin{figure}
   \psfig{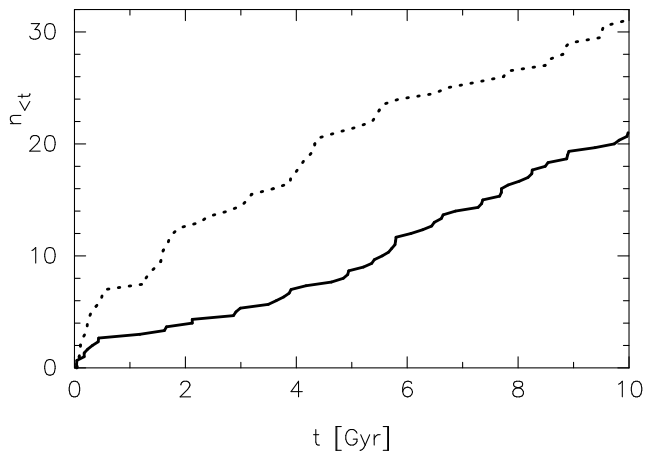}
   \caption[]{Cumulative distribution of merger times for (wd, wd)
   binaries for the compact clusters \#1 and \#2 (solid curve) and for
   the shallow clusters \#3 and \#4. These distributions were created
   from the cluster populations at an age of 100\,Myr.
   \label{fig:wdwdMergers}
   }
\end{figure}

Figure\,\ref{fig:wdwdMergers} shows the cumulative distributions of
merger times for those double degenerate binaries that merge within
10\,Gyr.  The double degenerates in the concentrated clusters (\#1 and
\#2) have an average merger time of $5.2\pm2.5$\,Gyr, compared to
$3.7\pm2.6$\,Gyr for those in the shallower clusters.  Thus, more
concentrated clusters tend to produce fewer double degenerate
binaries, with somewhat longer merger times.  This is probably related
to the tendency of the more concentrated clusters to contain more
binaries with shorter orbital period and more comparable component
masses.  Both effects are mediated by the higher interaction rates in
the denser clusters (see \S\,\ref{Sect:BlueStragglers}, where this
effect is discussed in relation to the formation of blue stragglers).

The reduced formation rate of merging white-dwarf binaries (the
potential sources of Type Ia supernovae) in the densest clusters
contrasts with the earlier findings of \cite{2002ApJ...571..830S}, who
found a ten-fold enhancement of the merger rate. However, the results
are hard to compare, as the initial conditions are quite different, as
are the effective time frames over which the simulations are
performed.

According to the binary population synthesis calculations of
\cite{2001A&A...375..890N}, the Galaxy contains about $1.1\times 10^8$
(wd, wd) binaries, giving rise to a merger rate of 0.011 per year.
Correcting this merger rate for initial conditions of the primordial
binary population adopted in our simulations, we find $\sim5\times
10^6$ (wd, wd) mergers during a Hubble time in the Galaxy, consistent
with the predicted merger rate of simulation models \#3 and \#4 if we
simply assume that all stars in the Galaxy formed in such clusters.
The distribution of mergers for these two models is presented as the
dotted line in Figure\,\ref{fig:wdwdMergers}, which has roughly equal
probability per unit time. For the densest clusters (simulations \#1
and \#2), however, the population of (wd, wd) binaries that merge in a
Hubble time is smaller by about one-third.  Binaries that favor blue
straggler formation (see \S\,\ref{Sect:BlueStragglers}) generally do
not produce short-period (wd, wd) binaries.
 
\begin{figure}
   \psfig{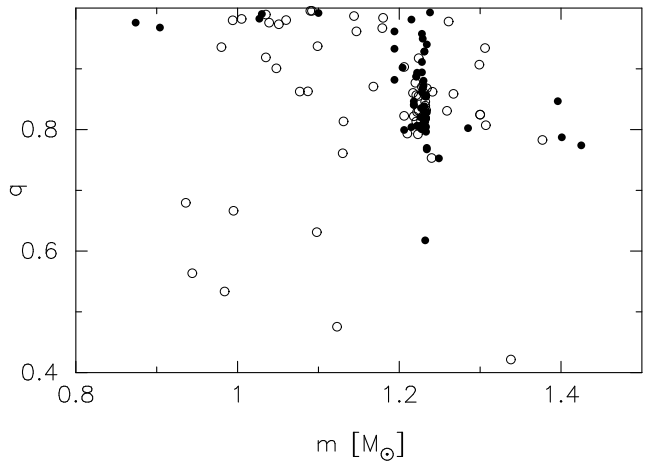}
   \caption[]{Distribution of primary mass versus mass ratio in (wd,
   wd) binaries at an age of 100\,Myr for the concentrated clusters
   \#1 and \#2 (bullets) and the shallow clusters \#3 and \#4 (open
   circles).
   \label{fig:MqwdwdMergers}
   }
\end{figure}

In the more concentrated clusters, the mean mass for the most massive
white dwarf in a binary is $\langle m\rangle = 1.23\pm0.03$\,\msun,
with a mass ratio of $\langle q \rangle = 0.87\pm0.07$ (see
Figure\,\ref{fig:MqwdwdMergers}).  For the shallower clusters, the
distributions in mass and mass ratio are much broader, but have
similar means: $\langle m\rangle = 1.21\pm0.09$\,\msun, $\langle q
\rangle = 0.84\pm0.08$.  
We note here that the potential merging white
dwarf binaries in the simulations of \cite{2002ApJ...571..830S} have
$\langle q \rangle = 0.69\pm 0.18$, considerably smaller than the
value found here.  However, differences in the cluster age have a
substantial effect on the final mass ratio, in the sense that, in
older star clusters like those of \cite{2002ApJ...571..830S}, the
secondary white dwarfs tend to be of lower mass.

The distributions of primary and secondary masses in double degenerate
binaries are presented in Figure\,\ref{fig:MqwdwdMergers}. The primary
mass white dwarfs in the denser clusters tend to cluster around
1.23\,\msun. The majority of these white dwarfs are formed following
stable mass transfer from an $\sim8-12\msun$ star.  This can only
happen in case A mass transfer \citep{kw67}, i.e., in a short-period
binary.  Denser clusters tend to produce more such systems, as
dynamical hardening of the binary reduces the orbital period, favoring
stable mass transfer.  The consequences of such reversed evolution
were discussed in \S\,\ref{Sect:CompactObjects}.

The larger number of dynamical encounters in the denser star clusters
leads to much narrower distributions in both primary mass and mass
ratio compared to the shallower clusters.  Interactions tend to put
the most massive stars into tighter binaries, and the combination of
tighter binaries and more massive companions, tends to cause mass
transfer in mass-transferring binaries to be more conservative.  As a
consequence, the resultant white dwarfs are more massive.  By the end
of our simulations (at 100\,Myr), only about 6\% of stars have evolved
into white dwarfs (see Figure\,\ref{fig:B_Stellarcontent}), whereas at
10\,Gyr every star will be a remnant (for our initial mass function),
so the population of binary white dwarfs is still expected to change
quite substantially.  However, the majority of massive white dwarfs in
relatively tight binaries, which are the most promising sources of
type Ia supernovae, and which are also easiest to observe by
gravitational wave observatories, form during the first 100\,Myr.

\section{Conclusions}

We have simulated star clusters with highly concentrated initial
density profiles and a wide range of initial relaxation times, from
birth to an age of about 100\,Myr.  Our initial conditions include
10\% hard primordial binaries, and the simulations incorporate the
effects of stellar and binary evolution and binary dynamics.

In this second paper on these simulations, we report on the time
variation of the structural and internal composition parameters
describing our model clusters, and compare our results directly to the
sample of relatively young and isolated star clusters in the Large
Magellanic Cloud.  On the basis of this comparison, we conclude that
the range of core radii and concentrations found in our simulated
clusters is consistent with observations of the LMC clusters, and we
argue that most of the LMC clusters are born with initial half-mass
relaxation times of 200\,Myr to 600\,Myr and high central
concentrations---$c \simeq 2.7$ (King parameter $\Wo \simeq 12$).  The
only clear exception to this is the star cluster R\,136 in the 30
Doradus region, which matches our simulation with an initial
relaxation time of about 80\,Myr.

Due to mass segregation, the mass function of intermediate-mass
main-sequence stars becomes as flat as $\alpha=-1.8$ in the central
part of the cluster (where the initial Salpeter mass function had
$\alpha=-2.35$).  In the outer regions, the mass function exponent is
as steep as $\alpha = -2.6$.  By the end of the simulations, at
100\,Myr, the overall cluster binary fraction is still $\sim 10$\,\%,
but in the core the fraction of binaries is somewhat higher ($\apgt
12$\%). By this time about 7\% of the single stars are remnants, and
their number is increasing gradually at a rate of about 0.1\% per Myr.

In our simulations a large number of blue stragglers are formed. At
any time, however, no more than 100--200 blue stragglers are visible
in the cluster.  The largest numbers of blue stragglers are formed in
the densest clusters.  The distribution of blue straggler masses
depends quite sensitively on the initial cluster density.  The densest
clusters tend to produce more massive, brighter and bluer blue
stragglers than less dense clusters.  The trends visible in our
simulations are consistent with observations of current globular
clusters.  A population of dormant blue stragglers is formed early in
the evolution of the cluster. They remain hidden on the main-sequence
until they emerge above the turn-off as the cluster ages.

The fraction of high-velocity stars of spectral type O and B is
considerably smaller than the fractions observed in the Galactic
field.  Our simulations, however, incorporate both of the effects
thought to be responsible for the acceleration of the observed OB
runaways: supernova in evolving binaries and gravitational slingshots
from multi-body scattering encounters.  The discrepancy with
observations of the numbers of OB runaways might conceivably be
explained by the initial binary fraction, which in our simulations is
only 10\%.

Shortly after formation, the cores of our simulated clusters become
quite rich in compact stars.  Up to an age of about 40\,Myr the
remnant population in cluster cores is dominated by stellar-mass black
holes; after that time white dwarfs take over. Neutron stars are
easily ejected from the clusters and there are only a few present at
any time.  The neutron star retention fraction is about 6--12\%,
whereas 50--70\% of black holes are retained.  Clusters with longer
relaxation times have smaller retention fractions.  Binaries
containing black holes with main-sequence companions outnumber those
containing a neutron star and a stellar companion.  We conclude that
these clusters may be relatively rich in x-ray binaries with a black
hole as accreting object, at least up to ages of a few hundred Myr.

Binaries containing two white dwarfs are quite common in our
simulations, and 20--30 have sufficiently small orbital periods that
gravitational radiation will bring the two white dwarfs into contact
within a Hubble time.  Interestingly, clusters with shorter relaxation
time produce systematically fewer white-dwarf binaries that will merge
within a Hubble time.

\section*{acknowledgments}

We are grateful to Douglas Heggie and Piet Hut for discussions. SPZ
and SMcM thank Tokyo University for the use of their GRAPE-6
hardware. The calculations were performed on the MODESTA GRAPE-6
systems in Amsterdam and the GRAPE-6 platform at Drexel University.
This research was supported in part by the Netherlands Organization
for Scientific Research (NWO grant No. 635.000.001 and 643.200.503),
the Netherlands Advanced School for Astronomy (NOVA), the Royal
Netherlands for Arts and Sciences (KNAW), the Leids Kerkhoven-Bosscha
fonds (LKBF) by NASA ATP grant NNG04GL50G.  Parts of the manuscript
were completed during a visit (by SPZ and SLWM) to the Kavli Institute
for Theoretical Physics at UC Santa Barbara, supported in part by the
National Science Foundation under Grant No. PHY99-07949.

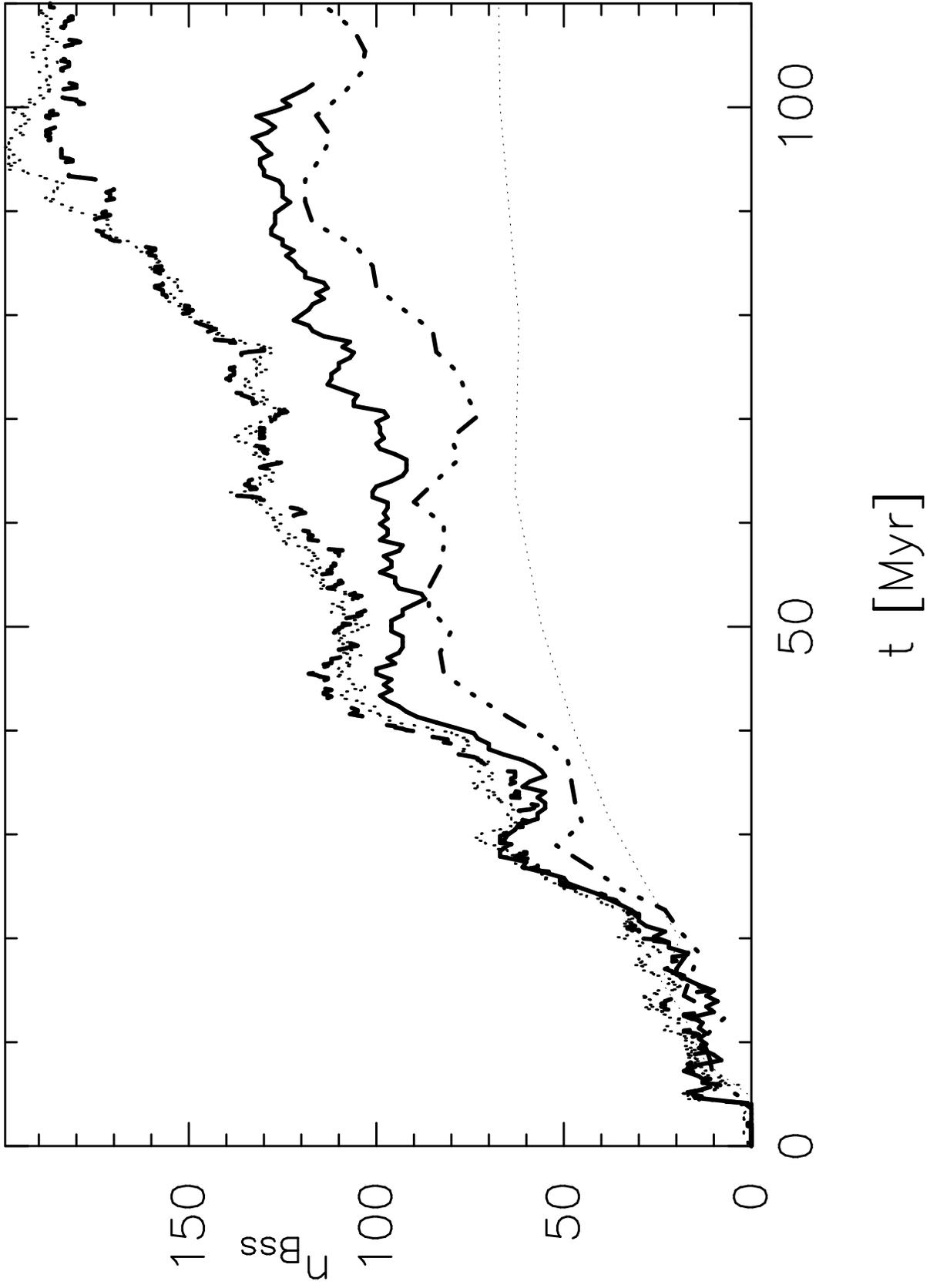
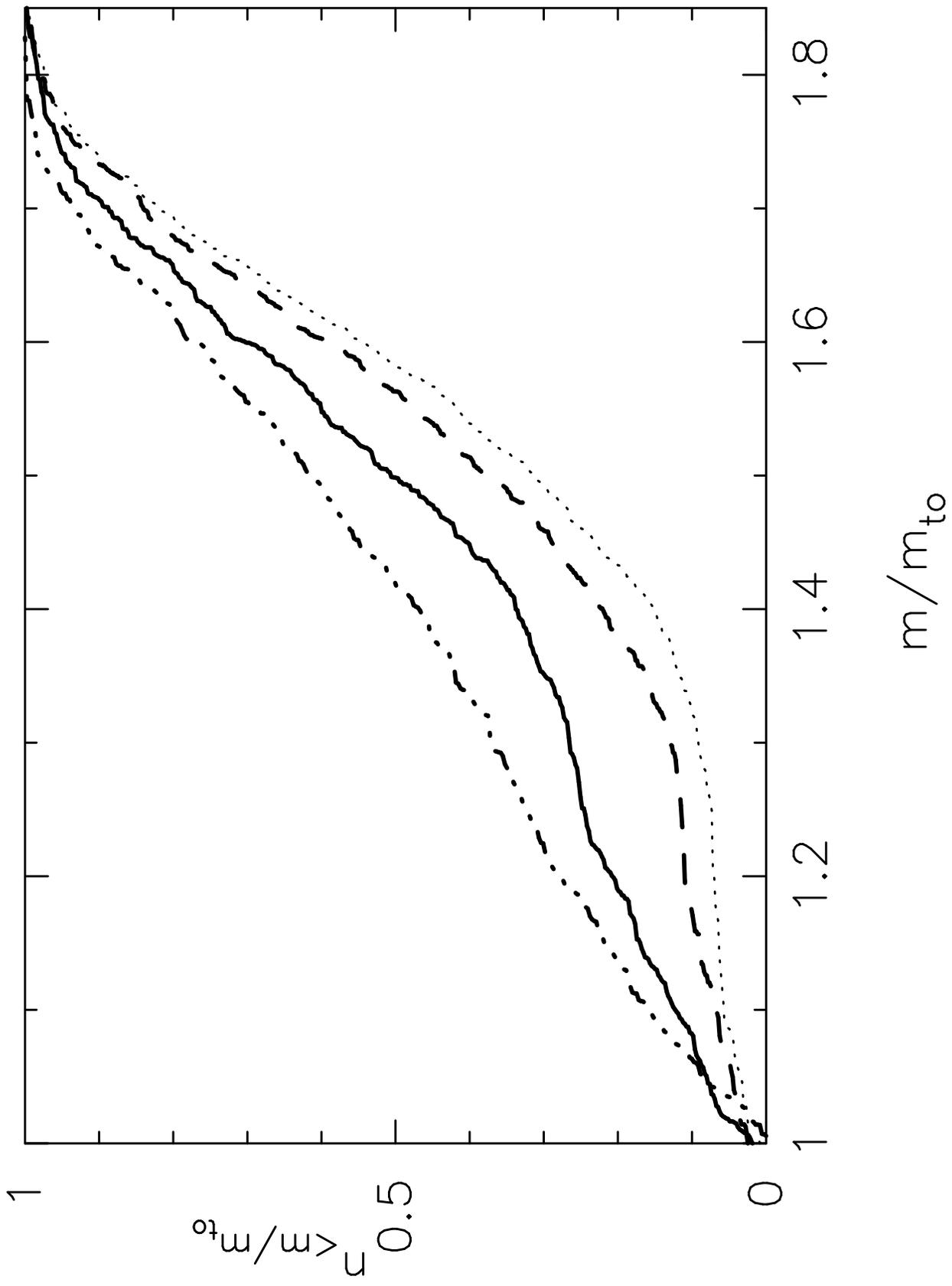
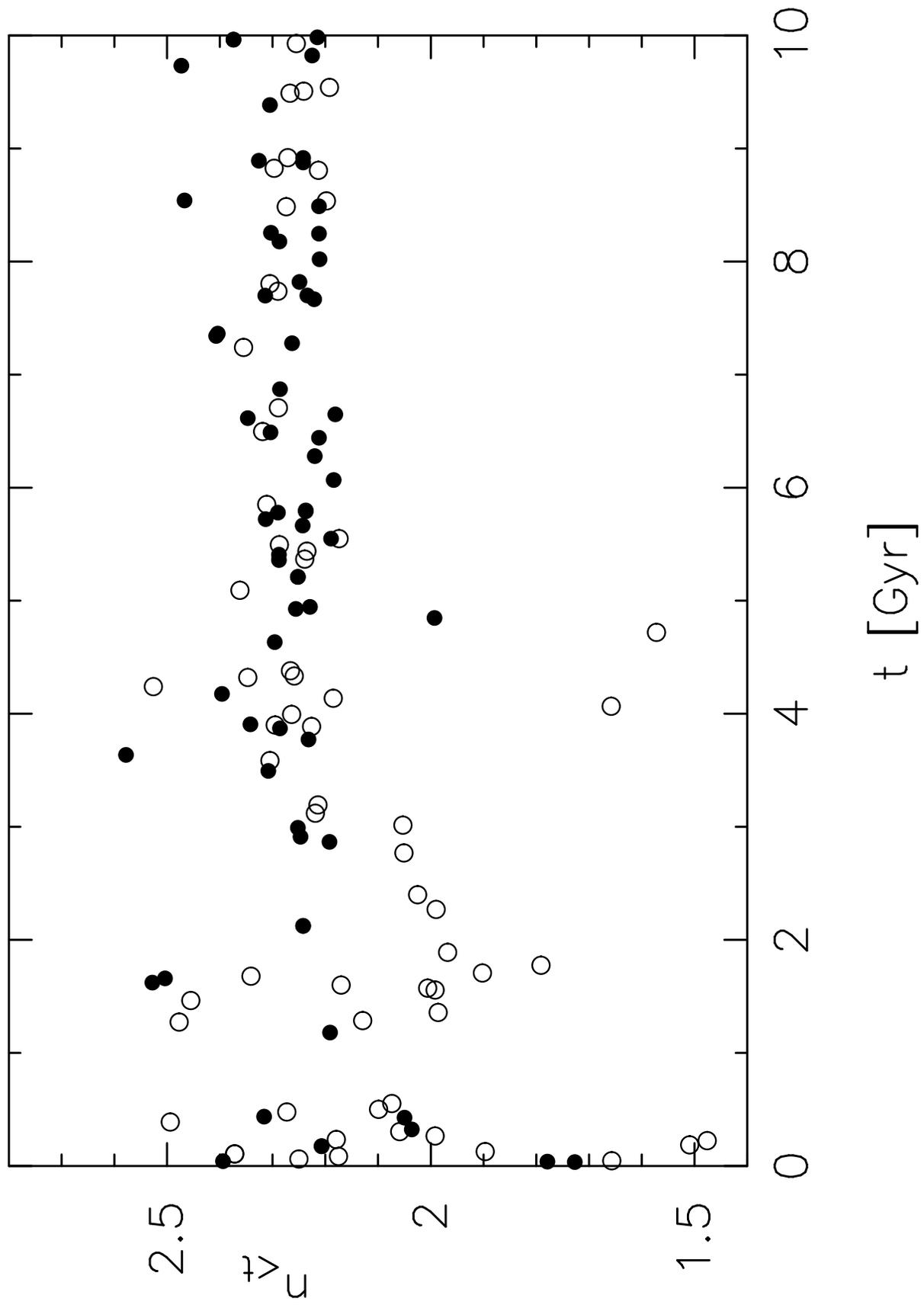
\end{document}